\documentclass[]{aastex62}

\usepackage[version=4]{mhchem}
\usepackage[utf8]{inputenc}

\graphicspath{{./}{figures/}}

\received{December 25, 2018}
\revised{February 8, 2019}
\accepted{February 11, 2019}

\submitjournal{ApJL}

\shorttitle{Titan Anion Chemistry}
\shortauthors{Dubois et al.}

\begin{document}

\title{Nitrogen-containing Anions and Tholin Growth in Titan's Ionosphere: Implications for Cassini CAPS-ELS Observations}

\correspondingauthor{David Dubois \\ Present addresses: NASA Ames Research Center, Space Science \& Astrobiology Division, Astrophysics Branch, Moffett Field, CA, USA, Bay Area Environmental Research Institute, Petaluma, CA, USA}
\email{david.f.dubois@nasa.gov}

\author[0000-0003-2769-2089]{David Dubois}
\affil{LATMOS/IPSL, UVSQ Université Paris-Saclay, UPMC Univ. Paris 06, CNRS, Guyancourt, France} 



\author[0000-0002-0596-6336]{Nathalie Carrasco}
\affil{LATMOS/IPSL, UVSQ Université Paris-Saclay, UPMC Univ. Paris 06, CNRS, Guyancourt, France} 

\author[0000-0003-4710-8943]{Jérémy Bourgalais}
\affil{LATMOS/IPSL, UVSQ Université Paris-Saclay, UPMC Univ. Paris 06, CNRS, Guyancourt, France} 

\author[0000-0002-2027-0943]{Ludovic Vettier}
\affil{LATMOS/IPSL, UVSQ Université Paris-Saclay, UPMC Univ. Paris 06, CNRS, Guyancourt, France}

\author[0000-0002-2015-4053]{Ravindra T. Desai}
\affil{Blackett Laboratory, Imperial College London, London, UK} 

\author{Anne Wellbrock}
\affil{Mullard Space Science Laboratory, University College London, Holmbury St. Mary, Dorking, Surrey RH5 6NT, UK}

\affil{Centre for Planetary Science at UCL/Birkbeck, London, Gower Street, London WC1E 6BT, UK} 

\author[0000-0002-6185-3125]{Andrew J. Coates}
\affil{Mullard Space Science Laboratory, University College London, Holmbury St. Mary, Dorking, Surrey RH5 6NT, UK}

\affil{Centre for Planetary Science at UCL/Birkbeck, London, Gower Street, London WC1E 6BT, UK}

\begin{abstract}

The Cassini Plasma Spectrometer (CAPS) Electron Spectrometer (ELS) instrument onboard Cassini revealed an unexpected abundance of negative ions above 950 km in Titan's ionosphere. \textit{In situ} measurements indicated the presence of negatively charged particles with mass-over-charge ratios up to 13,800 \textit{u/q}. At present, only a handful of anions have been characterized by photochemical models, consisting mainly of C$_n$H$^-$ carbon chain and C$_{n-1}$N$^-$ cyano compounds ($n=2-6$); their formation occurring essentially through proton abstraction from their parent neutral molecules. 
However, numerous other species have yet to be detected and identified. Considering the efficient anion growth leading to compounds of thousands of \textit{u/q}, it is necessary to better characterize the first light species. Here, we present new negative ion measurements with masses up to 200 \textit{u/q} obtained in an \ce{N2}:\ce{CH4} dusty plasma discharge reproducing analogous conditions to Titan's ionosphere. We perform a comparison with high altitude CAPS-ELS measurements near the top of Titan's ionosphere from the T18 encounter. The main observed peaks are in agreement with the observations. However, a number of other species (\textit{e.g.} \ce{CNN-}, \ce{CHNN-}) previously not considered suggests an abundance of N-bearing compounds, containing two or three nitrogen atoms, consistent with certain adjacent doubly-bonded nitrogen atoms found in tholins. These results suggest that an N-rich incorporation into tholins may follow mechanisms including anion chemistry, further highlighting the important role of negative ions in Titan's aerosol growth.

\end{abstract}



\section{Introduction}\ \label{sec:intro}

Ubiquitous anion chemistry present in circumstellar envelopes, clouds in the interstellar medium (ISM), planetary atmospheres and cometary comae, has for $\sim$4 Gya relied on the simplest anion found in space: \ce{H-}, formed through radiative electron attachment (REA). Necessary for the formation of \ce{H2} (see \cite{Millar2017} for a detailed review and references therein), \ce{H-} proceeds through \textit{e.g.} electron attachment (EA), REA or dissociative electron attachment (DEA) mechanisms.
To date, \ce{C8H-} is the largest carbon-chain anion detected in the ISM, i.e. the TMC-1 molecular cloud \citep{Brunken2007}. In planetary atmospheres, however, Saturn's largest moon Titan's ionosphere stands out as containing an abundance of unsuspected negative ions above 950 km (\cite{Coates2007,Waite2007,Agren2012}). 
Using CAPS-ELS (Cassini Plasma Spectrometer in electron spectrometer mode) measurements from 16 passes, \cite{Coates2007} revealed the presence of anions with mass-to-charge ratios up to 10,000, as well as latitude and altitude variations \citep{Coates2009}. 


The first photochemical model accounting for ionospheric negative ion species was subsequently investigated by \cite{Vuitton2009}, and focused on 11 low-mass negative ions. In spite of the lower mass resolution of CAPS, the isoelectronic series C$_{2n+1}$N$^-$, \textit{viz.} \ce{CN-}, \ce{C3N-} and \ce{C5N-} were predicted to be the three most abundant negative ions with a density peak near 1000 km \citep{Vuitton2009}. More recent analyses by \cite{Desai2017} additionally revealed the contribution of carbon-chain anions containing ethynyl groups (--C$_n$H, with $n = 2, 4, 6$), which could not be differentiated from the cyano-containing species. The location of peaks up to 165 \textit{u/q} were statistically constrained, and found an anti-correlation between the depletion of low-mass anions and the growth of larger ones $<1100$ km. These results suggested that low-mass anions may contribute to the growth of larger organic species with decreasing altitude, consistent with \ce{CN-}/\ce{C3N-} and \ce{C2H-}/\ce{C4H-} growth patterns. Nonetheless, only a handful of anions are presently characterized. An abundance of light and heavier species remains unidentified. 
Furthermore, \cite{Lavvas2013} studied aerosol growth by modeling the interaction of the aerosols in Titan's ionosphere through mass transfer between positively charged and negatively charged particles.\\


Currently, three photochemical modeling studies account for negative ions (\cite{Vuitton2009,Vuitton2018}; \cite{Dobrijevic2016} and \cite{Mukundan2018}). In the former studies, ion-pair formation, DEA, REA, DEA of supra-thermal electrons with nitriles such as HCN were found to be significant production pathways in the first steps of the anion chemistry.
 
 Recently, \cite{Mukundan2018} \citep[and since][]{Vuitton2018} revisited these density profiles for Titan's dayside T40 flyby with updated cross-sections and reaction rate coefficients. In agreement with prior studies, \ce{CN-} was found to be the dominant anion ( $\sim 5 \times 10^{-1} \ cm^{-3}$ at 1015 km) due to abundant HCN. However, the second most abundant anion was either \ce{H-} or \ce{C3N-}, with an inversion near the ionospheric peak \citep{Mukundan2018}.\\

Laboratory experiments have thus far mainly explored specific anion growth routes involving \ce{CN-} \citep{Zabka2012,Biennier2014} and \ce{C3N-} \citep{Bourgalais2016}. \cite{Wang2015,Wang2016} studied pathways involving negatively charged polyaromatic hydrocarbons (PAHs) with N and O atoms and \cite{Horvath2009} measured light negative ions in a high pressure point-to-plane \ce{N2}:\ce{CH4}:\ce{Ar} corona discharge. At present, more laboratory experiments are needed to investigate anion growth routes and composition in plasma conditions. Based on current and future experimental results, a better understanding of the anion gas phase precursors will lead to a clearer overall picture of how Titan's aerosols are formed in the upper atmosphere.\\

Here, we report mass spectrometry measurements performed in Titan-like simulations in order to assess light and intermediate anion precursors to tholin formation and subsequent growth. We have carried out negative ion measurements with mass detections of up to 200 \textit{u/q} in \ce{N2}:\ce{CH4} mixtures representative of Titan's ionospheric conditions. The spectra obtained are compared with CAPS ELS observations obtained during the T18 encounter in the 1--200 \textit{u/q} range. The role of the light and intermediate anion species and implications for tholin growth is also discussed.

\section{Methods} \label{sec:methods}\

In order to simulate analogous conditions found in Titan's ionosphere, we used the PAMPRE experiment \citep{Szopa2006}, a radio-frequency (RF) dusty plasma generated at 13.56 MHz. The RF source (30 W) delivers a capacitively-coupled plasma located between two electrodes, with a constant gas pressure of $\sim$1 mbar. The pressure here, higher than in Titan's ionosphere in order to compensate for the low collision frequency in the plasma and accelerate the gas product yield. The plasma was in a grid-free configuration \citep[see][]{Dubois2019}. The presence of a confining grid prevents the obstruction of the Quadrupole Mass Spectrometer (QMS).

Negative ion measurements were carried out with a Hiden Analytical EQP 200 QMS in contact with the plasma with an extraction potential of $+190$ V, through a 100 $\mu$m pinhole. The ions encountered a series of pre-optimized lenses and were guided through the ion optics, until they reached an energy filter and finally a Secondary Electron Multiplier (SEM) detector at $1800$ V. We used a 200 ms/u dwell time for mass analysis at 1 u steps over 1--200 \textit{u/q} while accumulating signal over 325 scans. Signal accumulation is necessary due to the low intensity of negative ions. Spectra are normalized over the highest intensity (at 26 \textit{u/q}) and only actual signal above the background electrical noise ($>10^{-3}$) is plotted.

\section{Results}
\subsection{Main anions and CAPS-ELS observations} \label{sec:CAPS}\



CAPS consisted of three sensors, one being ELS. The ELS was a top-hat electrostatic analyzer and initially designed to measure differential electron velocity distributions. Measurements of negative ions in Titan's cold ionosphere were obtained while the spacecraft was in the ram direction. Cassini being at supersonic velocities ($\sim$ 6 km/s), negative ions appear as sharp beam features seen by ELS \citep{Coates2007}. The CAPS actuator therefore swept across the field-of-view in the spacecraft's ram direction, resulting in multiple negative ion spectra during flybys. 
The observed ram energy-per-charge can be converted into mass-per-charge knowing the relative spacecraft speed and taking into account a spacecraft potential correction, assuming singly charged anions. This conversion theoretically enables identification of particles up to 150,000 \textit{u/q} \citep{Coates2007}, although to date $10^4$ \textit{u/q} represents the highest anion mass observed at Titan. The analyzer had an energy bandwidth $\Delta E/E = 16.7 \%$, corresponding to the energy resolution of the sensor. This roughly corresponds to a mass resolution of $\sim$ 6. In contrast, the ion mass spectrometer used in this study has a mass resolution of 100 at 100 \textit{u/q}, enabling us to distinguish more species in the laboratory.  \\

\begin{figure}
\centering
\includegraphics[scale=0.5]{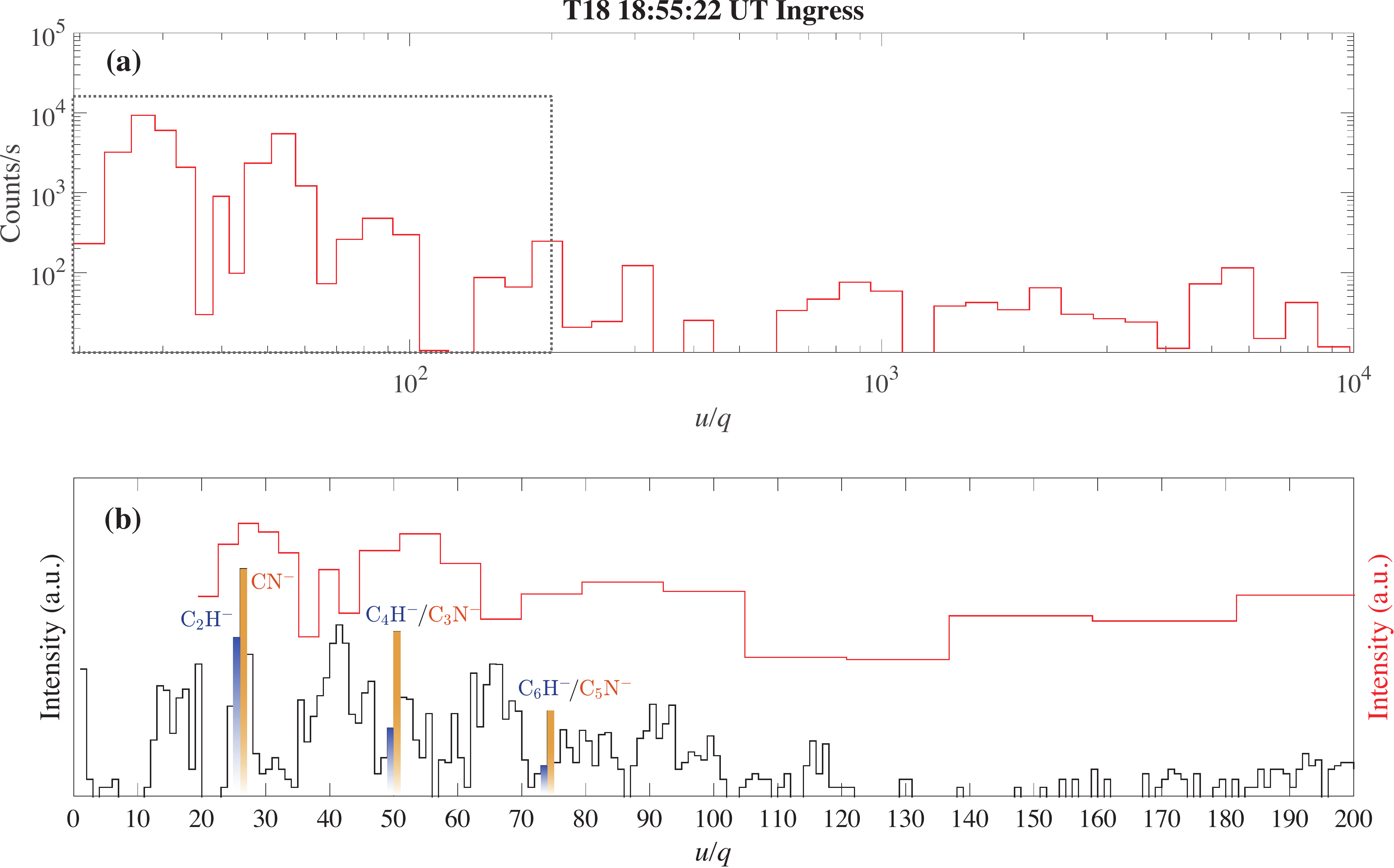}
\caption{(a) CAPS-ELS mass/charge ingress spectrum taken during the T18 ingress encounter at $\sim$ 1235 km. The dotted box represents the mass range plotted in the lower panel. (b) Comparison between the CAPS-ELS spectrum (red) and the registered experimental spectrum (black) taken in an \ce{N2}:\ce{CH4} 95:5 \% mixture (up to 200 \textit{u/q}), both shown in arbitrary intensity units. The bottom panel vertical axes are in log scale. Predicted species by \citet{Vuitton2009,Dobrijevic2016,Desai2017,Mukundan2018} are also shown. The blue marks correspond to the hydrocarbon anions \ce{C2H-}, \ce{C4H-} and \ce{C6H-} at 25, 49 and 73 \textit{u/q}, respectively. The N-bearing anions \ce{CN-}, \ce{C3N-} and \ce{C5N-} at 26, 50 and 74 \textit{u/q}, respectively, are labeled in orange. \label{fig1}}
\end{figure}

Figure \ref{fig1} shows a comparison between the experimental spectrum and CAPS-ELS measurements taken during the T18 encounter at $\sim$ 1235 km. This flyby occurred on September 23, 2006 over Titan's north polar region, entering on the nightside and exiting on the dayside, with a closest approach near the terminator. Background electron counts are removed by subtracting averaged counts from non-ram pointing anodes \citep{Coates2007,Wellbrock2013,Desai2017}.\\ 

The experimental spectrum was taken in an \ce{N2}:\ce{CH4} 95:5 \% plasma (black, bottom panel) with the major predicted anions. Approximately 8 grouings up to 120 \textit{u/q} in the experimental spectrum are found (Table \ref{Table1}), with main peaks alternatively separated by 15 and 9 u from 26 to 74 \textit{u/q}. 
Although few ions are detected between 120 and 150 \textit{u/q}, a secondary increase is observed starting at 155 \textit{u/q} up to the mass detection limit 200 \textit{u/q}. The most intense ion at 26 \textit{u/q} is attributed to \ce{CN-} \citep{Vuitton2009}. The other cyano group anions \ce{C3N-} and \ce{C5N-} are detected at 50 and 74 \textit{u/q}, respectively. Early work by \cite{Vuitton2009} suggested that \ce{CN-} was mainly formed via DEA of HCN resulting from supra-thermal electron impact. More recently, however, \cite{Mukundan2018} and \cite{Vuitton2018} used updated DEA cross-sections of HCN, resulting in proton transfer between \ce{H-} and HCN (Reaction \ref{CN- formation}) as the major pathway to \ce{CN-} formation. \ce{H-} is mainly produced through DEA of \ce{CH4} \citep{Mukundan2018}. 

\begin{equation}
\label{CN- formation}
\ce{H- + HCN \longrightarrow CN- + H2}
\end{equation}\\

The main pathways for heavier species \citep{Biennier2014,Dobrijevic2016,Mukundan2018,Vuitton2018} are given below. \cite{Mukundan2018} find REA as the main production pathway to \ce{C3N-} due to differences in the neutral profiles used in both models.

\begin{equation}
\label{C3N- formation}
\ce{CN- + HC3N \longrightarrow C3N- + HCN}
\end{equation}

\begin{equation}
\label{C5N- formation}
\ce{CN- + HC5N \longrightarrow C5N- + HCN}
\end{equation}\\

These mass patterns are consistent with the ELS distributions $>$ 20 \textit{u/q} (Figure \ref{fig1}, bottom panel) notably at 26 \textit{u/q}. At 35-40 \textit{u/q}, we note an intermediate peak not reported previously in the ELS data, matching our mass group 3 surprisingly well. This further peak is only discernable intermittently at high Titan altitudes and is not observed during all encounters (e.g., it is absent in the T40 spectrum taken at 1244 km in \cite{Desai2017}). The 50 \textit{u/q} peak is also consistent with the 45-70 \textit{u/q} group present in the ELS spectrum. For larger mass species, it is noteworthy to point out that anions could be present at nearly every mass per unit charge of the spectrum given the continuous distribution of ELS over this mass range. At higher masses ($>$70 \textit{u/q}), the laboratory spectrum further seems to match the ELS data at 70-90 and 110-120 \textit{u/q}, which were characterized by \cite{Desai2017}. A further increase in intensity starting at 140-150 \textit{u/q} is seen in both spectra, although the low ion signal currently renders the interpretation difficult.\\





\begin{table}[]
\caption{Broad mass groups from 1--200 \textit{u/q} defined by \cite{Wellbrock2013} (updated from \cite{Coates2007}) obtained from all first 34 encounters. These mass groups are qualitatively compared with groups defined from the experimental spectrum degraded to the CAPS resolution and the experimental spectrum from Figure \ref{fig1} at much higher resolution.}
\begin{tabular}{cc|cc|cc}
\hline
\multicolumn{2}{c|}{\cite{Wellbrock2013}}   & \multicolumn{2}{c|}{\begin{tabular}[c]{@{}c@{}}This study\\ mass groups (CAPS resolution)\end{tabular}} & \multicolumn{2}{c}{\begin{tabular}[c]{@{}c@{}}This study\\ mass groups (experimental resolution)\end{tabular}} \\ \hline 
mass groups          & \textit{u/q}                   & mass groups                                        & \textit{u/q}                                                & mass groups                                              & \textit{u/q}                                                 \\ \hline \hline 
1                    & 12-30                 & 1                                                  & 13-31                                              & 1                                                        & 11-18                                               \\
2                    & 30-55                 & 2                                                  & 37-61                                              & 2                                                        & 23-34                                               \\
3                    & 55-95                 & 3                                                  & 67-102                                             & 3                                                        & 35-46                                               \\
4                    & 95-130                & 4                                                  & 115-121                                            & 4                                                        & 47-55                                               \\
5                    & 130-190               & 5                                                  & 169-200                                            & 5                                                        & 56-71                                               \\
6                    & 190-625               & \multicolumn{1}{l}{}                               & \multicolumn{1}{l|}{}                              & 6                                                        & 72-85                                               \\
7                    & 625+                  & \multicolumn{1}{l}{}                               & \multicolumn{1}{l|}{}                              & 7                                                        & 87-101                                              \\
\multicolumn{1}{l}{} & \multicolumn{1}{l|}{} & \multicolumn{1}{l}{}                               & \multicolumn{1}{l|}{}                              & 8                                                        & 104-122                                             \\
\multicolumn{1}{l}{} & \multicolumn{1}{l|}{} & \multicolumn{1}{l}{}                               & \multicolumn{1}{l|}{}                              & 9                                                        & 150+     \\                             
\hline
\label{Table1}
\end{tabular}
\end{table}


\subsection{Spectra analysis of light and intermediate N-rich anions}\ \label{sec:analysis}


\begin{figure}
\plotone{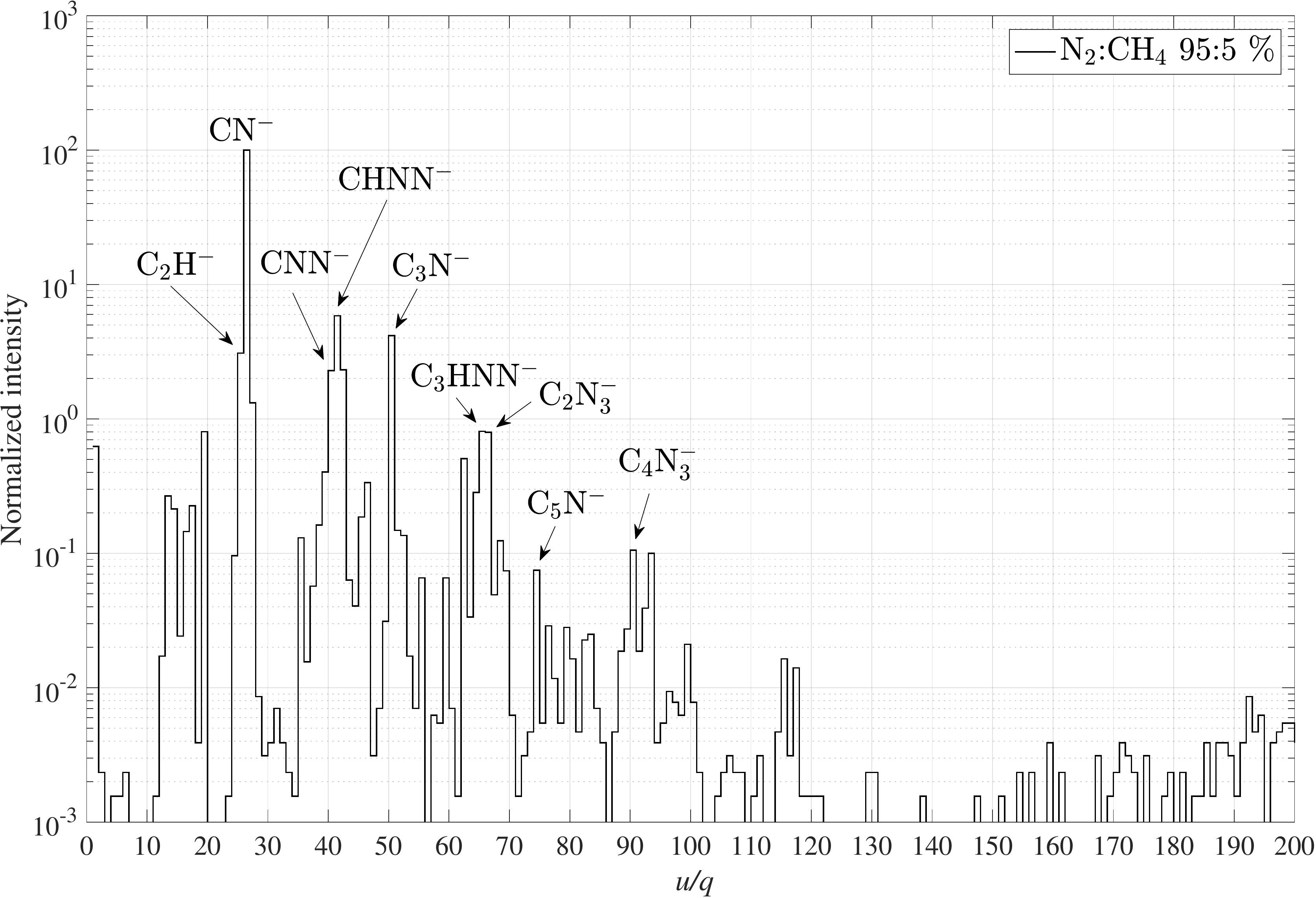}
\caption{Registered mass spectrum of anions detected in an \ce{N2}:\ce{CH4} 95:5 \% plasma discharge, over 1--200 \textit{u/q}, accumulated over 325 scans. 
Tentative attributions are shown for the major peaks.  \label{fig2}}
\end{figure}


Figure \ref{fig2} shows the same experimental spectrum as in Figure \ref{fig1} (bottom panel) with attributions of species $<$ 100 \textit{u/q}.
Unexpected intense peaks stand out at 41, 65 and 66 \textit{u/q}. \cite{Horvath2009}, which used a point-to-plane corona discharge at higher pressure, putatively assigned the peaks at 41 and 65 \textit{u/q} to the diazirinyl \ce{CHNN-} and dicyanomethanide \ce{C3HNN-} anions, respectively. Other species which could correspond to these peaks according to the NIST are \ce{C2H3N-}, \ce{C3H5-} and \ce{C5H5-}. Comparisons with tholin signatures and gas phase pathways are detailed hereafter, and confirm the attributions by \cite{Horvath2009}. Note the presence of a peak at 19 \textit{u/q} which likely corresponds to a fluorine \ce{F-} contamination. The presence of \ce{F-} is presumably due to polytetrafluoroethylene (PTFE) from the temperature probe sheaths located near the polarized electrode in the reactor. This signal appears later over the accumulation, indicating its contamination does not affect the overall signal of the spectrum. Also in this mass group, NH3 is inferred by the attribution of \ce{NH2-} at 16 \textit{u/q} \citep{Horvath2010b,Horvath2011}.\\

\ce{CHNN-} can either be in a cyclic four-$\pi$ electron conformer or as an open biradical isomer \citep{Gordon1995}. These diazo isomer compounds are stable and can contain nitrogen-nitrogen double bonds \citep{Kroeker1990,Gordon1995}.\

Current photochemical models largely account for nitrile incorporation of products containing only one nitrogen atom. Diazo and triazo species however, are not taken into account and their implications on the gas phase chemistry are poorly understood with respect to the nitrogen-rich tholins. Two previous studies by \cite{Gautier2011} and \cite{Cunha2016} on the gas phase and solid tholin material produced in the PAMPRE reactor can be clues to this conundrum. \cite{Gautier2011} first identified \ce{C4H3N5} (Tetrazolo[1,5-b]pyridazine) as the main aromatic gas phase product using gas chromatography. This heteroaromatic compound contains multiple single or double bonded nitrogen atoms. Furthermore, \cite{Cunha2016} revealed the presence of heteroaromatic structures present in tholins such as triazole rings (\ce{C2H3N3}). This compound, along with the cyclic diazirine \ce{c-CH2N2} or cyanamide \ce{H2N-CN}, were shown to be strong candidates for the observed \ce{N=N} patterns. These isomers may indeed contain doubly-bonded nitrogen atoms, which may account for the direct incorporation of \ce{N2} into tholins. Such N-rich patterns can be directly linked with the anions assigned here. The diazo and triazo radicals attributed in Figure \ref{fig2} could indeed serve as important intermediates for the heterogeneous chemistry observed in tholins. Nitrogen incorporation into tholins may thus be facilitated by the potential presence of these nitrogen-rich (diazo and triazo) compounds in the gas phase. 

Furthermore, the putative presence of intermediate species such as \ce{CHNN-} (41 \textit{u/q}) could contribute more or less significantly to the overall 35-60 \textit{u/q} mass grouping to which \cite{Desai2017} found a nominal mass center of 49 \textit{u/q}, within the $2\sigma$ uncertainty, near closest approach. Non-attributed intermediate compounds formed seen here at high altitudes could potentially still be present in the ELS continuum at lower altitudes. Therefore, such a mass-over-charge shift to slightly lower values could indicate the presence of these intermediate compounds falling within the fitted distribution. \\

In spite of limited knowledge at present on diazo anion chemistry, we propose below mechanisms describing the putative presence of \ce{CNN-} and \ce{CHNN-} (40 and 41 \textit{u/q}) or their possible isomers \ce{NHCN-} and \ce{NCN-}. \ce{CH2CN-} could also be a candidate at 40 \textit{u/q}.\\





A candidate molecule within this mass range potentially available for the initiation of further pathways is \ce{CH2N2} \citep[\textit{e.g.}][]{He2014,Cunha2016}, for which three isomeric forms are considered here: diazomethane \ce{CH2N2}, cyclic diazirine \ce{c-CH2N2} and cyanamide \ce{H2N-CN}. \ce{CH2N2} was previously excluded from the composition of tholins by \cite{Cunha2016}, while both cyclic diazirine and cyanamide have been identified among several nitrogenous prebiotic organic compounds present in tholins \citep{He2013,He2014,Cunha2016}. Formation of linear \ce{CH2N2} (diazomethane) following \ce{1CH2 + N2 \longrightarrow CH2N2} was shown by \cite{Braun1970} to be significant only for high pressures not relevant to our experimental setup nor to Titan's upper atmosphere conditions. \ce{HCN} and \ce{NH3} alone were suggested to contribute to the formation of cyanamide \ce{H2N-CN} (and its dimer dicyanodiamide \ce{C2H4N4}) according to Reaction \ref{cyanamide He and Smith 2013} \citep{He2013}. The cyanamide isomer can also go through a number of low energy EA or DEA reactions (Reactions \ref{DEA Tanzer}, \ref{DEA Tanzer 2} and \cite{Tanzer2015}). 

\begin{equation}
\label{cyanamide He and Smith 2013}
\ce{HCN + NH3 \longrightarrow H2N-CN + H2}
\end{equation}

\begin{equation}
\label{DEA Tanzer}
\ce{H2N-CN + e- \longrightarrow NHCN- + H \ (41 u/q)}
\end{equation}

\begin{equation}
\label{DEA Tanzer 2}
\ce{H2N-CN + e- \longrightarrow CNN- + H2 \ (40 u/q)}
\end{equation}\\





Similar routes from the parent neutral diazirine \ce{c-CH2N2} may also occur (Reactions \ref{diazirine1} and \ref{diazirine2}).

\begin{equation}
\label{diazirine1}
\ce{c-CH2N2 + e- \longrightarrow c-CH2N2- \ (42 u/q)}
\end{equation}

\begin{equation}
\label{diazirine2}
\ce{c-CH2N2 + e- \longrightarrow CNN- + H2 \ (40 u/q)}
\end{equation}\\

In group 5, the other diazo anion \ce{C3HNN-} (dicyanomethanide) attributed to the peak at 65 \textit{u/q}, following \cite{Horvath2009}, has an intensity of an order of magnitude greater than \ce{C5N-}. It is accompanied by a second intense peak at 66 \textit{u/q}. This mass can correspond to two anions, \ce{C4H4N-} and \ce{C2N3-} (the pyrrolide anion and dicyanamide, respectively). \ce{C4H4N-} was tentatively attributed by \cite{Horvath2009}, whose parent neutral is pyrrole \ce{C4H5N}. Given the previous detections of pyrrole and other heterocyclic aromatic compounds in this and other similar plasma discharge experiments as well as in tholin material \citep[\textit{e.g.}][]{McGuigan2006,Gautier2011,He2014,Mahjoub2016}, such an ion can be a candidate for 66 \textit{u/q}. However, \cite{Carrasco2009} carried out analyses of the soluble fraction of tholins and measured charged ionic species by collision induced dissociation (see also \cite{Somogyi2012}). They found recurrent ionic fragments at 66 \textit{u/q} which, given its fragmentation patterns observed in negative ion mass spectrometry, revealed dicyanamide \ce{C2N3-} to be a better candidate at this mass-to-charge ratio. Therefore, we attribute the gas phase N-containing carbanion precursor \ce{C2N3-} as a likely compound at the 66 \textit{u/q} peak. Interestingly, this compound has a v-shaped bent conformation (C$_{2V}$) with a permanent dipole moment, and was found to be unreactive and extremely stable, potentially existing in cold regions of the ISM \citep{Yang2011,Nichols2016}.
Furthermore, it was shown by \cite{Carrasco2009} and \cite{Somogyi2012} that \ce{C2N3-} was part of a fragmentation pattern involving HCN, \ce{C2H2} and \ce{NH2CN} formal additions. The attribution here of \ce{C2N3-} and its relative intensity in the gas phase sustains its important role as a precursor "seed" as proposed in the aforementioned studies. This N-containing carbanion may indeed play an important role in the polymerization growth of tholins and large gas phase N-rich precursors \citep{Carrasco2009,Somogyi2012}. 

Finally, \ce{C2N3-} (along with \ce{C4N3-}) has no substantial reactivity with protons, and its stability in hydrogen and N-containing carbanion-rich environments may make it a viable candidate as a gas phase precursor \citep{Bierbaum2011}. Given the bent and polar structure of \ce{C2N3-}, future rotational spectrum analyses of this compound should help in its potential detection in planetary atmospheres and the ISM.\



\section{Summary}\ \label{sec:Summary}

Our current understanding of the growth pathways of anions and their role as precursors to haze macromolecules Titan's upper atmosphere is limited. Measurements performed by CAPS-ELS revealed unidentified compounds up to 13,800 \textit{u/q} in the ionosphere. State-of-the-art photochemical models have attributed only a few of the first ions as carbon chain anions \ce{CN-}/\ce{C3N-} and \ce{C2H-}/\ce{C4H-}. In order to further investigate the anion chemistry and composition responsible for this efficient growth, we conducted our first direct anion measurements with \textit{u/q} up to 200 in the PAMPRE dusty plasma discharge. Spectra were taken in an \ce{N2}:\ce{CH4} 95:5 \% gas mixture, representative of Titan's ionospheric conditions. The broad mass groupings obtained in the laboratory spectra are consistent with recent ELS data analyses. These results are obtained at a higher resolution than the ELS data and reveal new and abundant N-rich compounds. It is therefore plausible that some of these compounds may be present in Titan's ionosphere, given the difficulty in interpreting the continuous ELS data at higher masses. Notable peaks at 26, 41, 50, 65 and 66 \textit{u/q} are attributed to \ce{CN-}, \ce{CHNN-}, \ce{C2N3-}, \ce{C3HNN-}, \ce{C2N3-}, respectively. In particular, the diazo and triazo compounds appear to be important N-rich anion precursors to tholin formation. Such a nitrogen incorporation into Titan tholins is compatible with \ce{N=N} patterns seen in tholins. Anion chemistry may therefore be a privileged route for N-bearing carbanions to form large compounds and ultimately to be incorporated into the compositional structure of tholins. Lastly, these results also call for further experimental, theoretical and observational studies working comprehensively to decipher the unsuspected and complex anion chemistry of Titan's ionosphere.



\acknowledgments

The authors are thankful to two anonymous reviewers who participated in the improvement of the manuscript. We also thank V. Mukundan, Dr. E. Sciamma-O'Brien and Dr. J. Westlake for insightful discussions. 
D.D., N.C., J.B. and L.V. acknowledge financial support by the European Research Council Starting Grant PrimChem, grant agreement no. 636829. RTD was supported by NERC grant NE/P017347/1. A.J.C. and A.W. acknowledge support from the STFC consolidated grant to UCL-MSSL ST/N000722/1, and ESA and UKSA for CAPS ELS operations support.

\end{document}